\newif\ifsingle

\ifsingle
\documentclass[11pt,draftclsnofoot, onecolumn]{IEEEtran}		
\else		
\documentclass[10pt,final, twocolumn]{IEEEtran}
\fi

\usepackage{times}
\usepackage{amsmath,dsfont}
\usepackage{amssymb,amsthm}
\usepackage{epsfig,verbatim}
\usepackage{caption}
\usepackage{subcaption}
\usepackage{setspace}
\usepackage{color}
\usepackage{cite}
\usepackage{epstopdf}
\usepackage{graphicx}
\usepackage{accents}
\usepackage{acronym}
\usepackage[bookmarks,colorlinks]{hyperref}
\usepackage{booktabs}
\usepackage{mathtools} 
\usepackage{enumitem}
\usepackage{multirow}
\usepackage{soul}

\ifsingle

\else

\fi 
\acrodef{adc}[ADC]{analog-to-digital convertor}
\acrodef{cs}[CS]{compressed sensing}
\acrodef{cnn}[CNN]{convolutional neural network} 
\acrodef{dnn}[DNN]{deep neural network} 
\acrodef{csi}[CSI]{channel state information}
\acrodef{map}[MAP]{maximum a-posteriori probability}
\acrodef{snr}[SNR]{signal-to-noise ratio}
\acrodef{bs}[BS]{base station} 
\acrodef{iot}[IoT]{Internet of Things}
\acrodef{mimo}[MIMO]{multiple-input multiple-output}
\acrodef{mse}[MSE]{mean-squared error}
\acrodef{pdf}[PDF]{probability density function}
\acrodef{rv}[RV]{random variable}
\acrodef{fec}[FEC]{forward error correction} 
\acrodef{lti}[LTI]{linear time-invariant}
\acrodef{wss}[WSS]{wide-sense stationary}
\acrodef{psd}[PSD]{power spectral density}
\acrodef{ser}[SER]{symbol error rate} 
\acrodef{ber}[BER]{bit error rate} 
\acrodef{sgd}[SGD]{stochastic gradient descent} 
\acrodef{isi}[ISI]{intersymbol interference}  
\acrodef{awgn}[AWGN]{additive white Gaussian noise} 
\acrodef{ut}[UT]{user terminal} 
\acrodef{mmw}[mmWave]{millimeter wave}
\acrodef{ai}[AI]{artificial intelligence} 
\acrodef{vqvae}[VQ-VAE]{vector quantized variational autoencoder}

\IEEEoverridecommandlockouts

\title{Collaborative Inference for AI-Empowered\\ IoT Devices}

\author{  
	\IEEEauthorblockN{Nir Shlezinger,~\IEEEmembership{Member,~IEEE}  and Ivan V. Baji\'{c},~\IEEEmembership{Senior Member,~IEEE}\\
	}  
	\thanks{
		 N. Shlezinger is with the School of Electrical and Computer Engineering, Ben-Gurion University of the Negev, Beer-Sheva, Israel (e-mail: nirshl@bgu.ac.il).
		I. V. Baji\'{c} is with the School of Engineering Science, Simon Fraser University, Burnaby, BC V5A 1S6, Canada (e-mail:  ibajic@ensc.sfu.ca). }
} 
 
\begin{document}
	
	\maketitle
 	\pagestyle{plain}  
\thispagestyle{plain}

\begin{abstract}
Artificial intelligence (AI) technologies, and particularly deep learning systems, are traditionally the domain of large-scale cloud servers, which have access to high computational and energy resources. Nonetheless, in Internet-of-Things (IoT) networks, the interface with the real-world is carried out using edge devices that are limited in hardware and can communicate. The conventional approach to provide AI processing to data collected by edge devices involves sending samples to the cloud, at the cost of latency, communication, connectivity, and privacy concerns. Consequently, recent years have witnessed a growing interest in enabling AI-aided inference on edge devices by leveraging their communication capabilities to establish {\em collaborative inference}. 
This article reviews candidate strategies for facilitating the transition of AI to IoT devices via collaboration. We identify the need to operate in different mobility and connectivity constraints as a motivating factor to consider multiple schemes, which can be roughly divided into methods where inference is done remotely, i.e., on the cloud, and those that infer on the edge. We identify the key characteristics of each strategy in terms of inference accuracy, communication latency, privacy, and connectivity requirements, providing a systematic comparison between existing approaches. We conclude by presenting future research challenges and opportunities arising from the concept of collaborative inference.
\end{abstract}

\section{Introduction} 
The philosophical idea of \ac{ai}, dating back multiple decades, 
is nowadays evolving into reality. Deep learning is demonstrating unprecedented success in a broad range of applications: \acp{dnn} surpass human ability in classifying images; reinforcement learning allows computer programs to  defeat human experts in challenging games; generative models create images of fake people, which appear indistinguishable from true ones. The successful combination of deep learning with the expected proliferation of smart edge devices, and particularly \ac{iot} devices, is expected to bring \ac{ai} to many aspects of our lives, ranging from  intelligent wearable sensors to self-driving vehicle and smart manufacturing systems.

Deep learning, which is the key enabler technology for \ac{ai}, relies on highly-parameterized models, trained using massive volumes of data. Consequently, deep learning is traditionally the domain of large scale computer servers, which have the computational resources and the ability to aggregate the data required to store, train, and apply \acp{dnn}. However, edge devices do not share these computational and storage 
resources, 
making the transition of \ac{ai} from the domain of powerful servers to distributed and computationally-limited edge devices a challenging task~\cite{chen2019deep}. 
 
The challenges associated with using \acp{dnn} on \ac{iot} edge devices can be divided according to the main machine learning tasks: training and inference. The challenges related to the former stem from the fact that edge devices have access only to a fraction of the data that can be aggregated by centralized \ac{ai} systems, yet sharing this data with a centralized server may give rise to privacy concerns and limit the ability to train personalized models. Schemes for enabling learning on the edge are widely studied in the literature, with arguably the most common approach being the federated learning paradigm~\cite{gafni2021federated}, where multiple devices collaborate during training in a centrally orchestrated fashion. 

Even when one has access to a trained \ac{ai} model, having it utilized by \ac{iot} devices gives rise to many different challenges. Most notably, \acp{dnn} 
 are often comprised of millions and even billions of parameters. Hence, hardware-limited \ac{iot} devices may be unable to merely apply such trained \acp{dnn} due to storage, energy, and computational considerations. A common practice is to have the edge device communicate their measurements to a powerful centralized server for inference. Yet, this strategy induces delay, gives rise to privacy concerns, imposes a notable burden on the server, and limits \ac{ai}-aided inference to settings where reliable communications with the server is attainable. These challenges can become limiting factors for emerging applications such as, e.g., \ac{ai}-empowered wearable devices. Consequently, recent years have witnessed a growing interest in leveraging collaboration for facilitating edge inference, with the proposal of various different techniques \cite{huang2022toward,malka2022decentralized,
 mao2017survey}, motivating the unified overview of these methods.
 
 In this article we systematically review candidate approaches for enabling \ac{ai}-aided inference on \ac{iot} devices. 
 While the successful transition of \ac{ai} to \ac{iot} devices is likely to rely on developments in both hardware as well as signal processing and algorithmic techniques, our focus is on the latter.
 We commence with discussing the diverse use-cases for \ac{iot} inference, reviewing their associated  characteristics and positioning them in the context of the conventional paradigms of edge vs cloud computing \cite{mahmud2018fog}. We particularly divide these settings into static scenarios, as arise in, e.g., smart manufacturing systems, and dynamic mobile scenarios, relevant to, e.g., wearable \ac{iot} devices and vehicular systems. This division reveals the broad range of requirements and the need for  diverse collaborative strategies for  \ac{ai}-aided edge inference.
  
  Next, we categorize existing and emerging approaches for  \ac{ai}-aided inference into two main strategies:  $1)$ inference on a central system, either  a dedicated edge server or a remote cloud server, with collaboration potentially used to relieve latency, congestion, and privacy concerns; $2)$ on-device inference, either in a purely decentralized or in a centrally orchestrated manner, where collaboration can improve performance with different levels of compensation for latency, privacy, and flexibility. We provide 
  comparisons between these approaches, capitalizing on their individual pros and cons in light of the identified families of expected use-cases. We conclude by discussing the road ahead and key research challenges that are yet to be explored to fully harness the potential of collaboration techniques for facilitating high-performance low-latency \ac{iot} inference.


\begin{figure}
    \centering
    \includegraphics[width=\columnwidth]{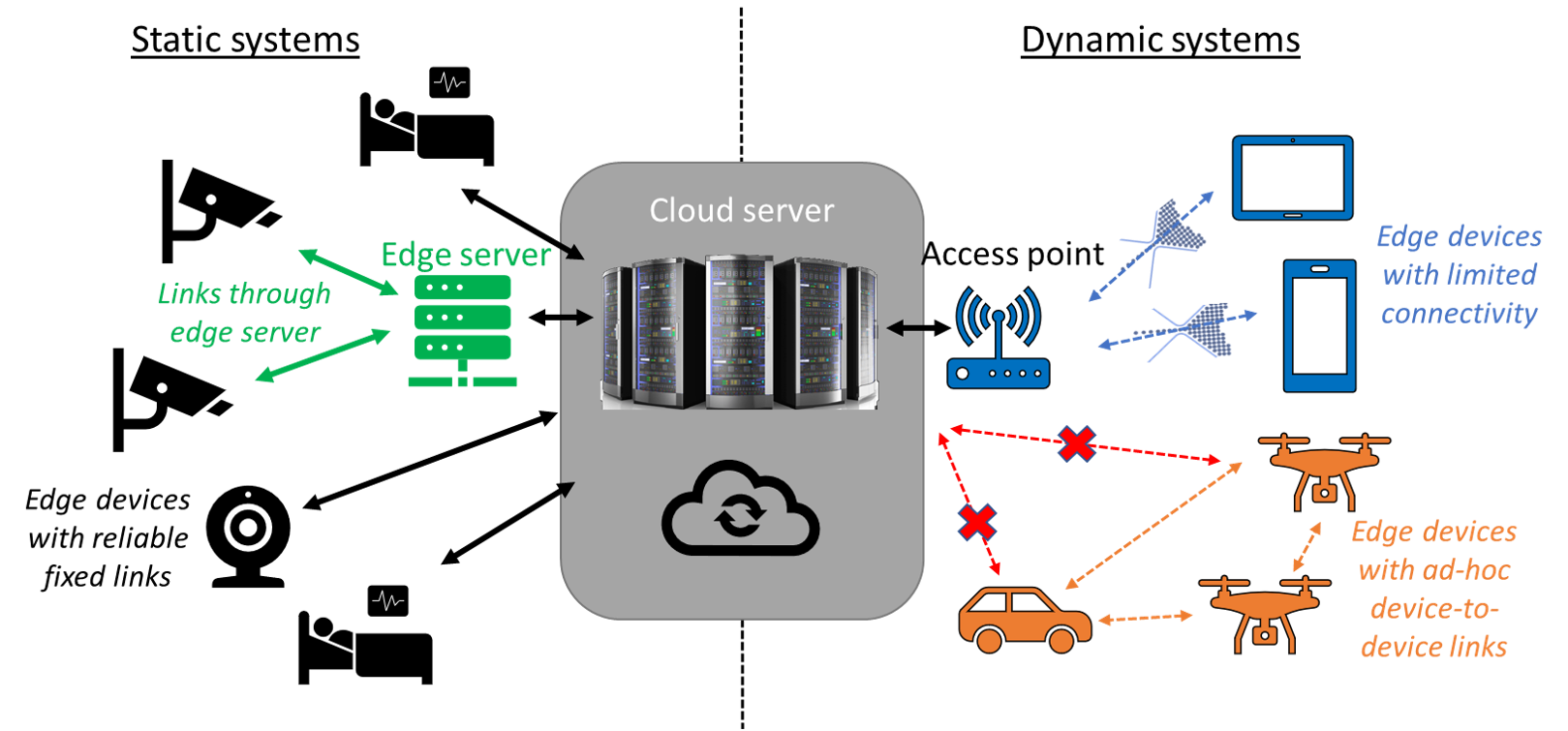}
    \caption{Illustration of \ac{iot} inference in static settings (left), where the devices have constant guaranteed links with the cloud server, versus dynamic settings (right), in which the devices are mobile and operate in diverse levels of connectivity.}
    \label{fig:DynVsStatic1}
\end{figure}

\section{AI-Empowered IoT Inference}
\label{sec:Inference}
The term \ac{iot} encompasses a broad range of devices that possess the ability to sense and communicate. Consequently, inference tasks associated with \ac{iot} devices  span a wide scope of diverse use-cases. Diversity in \ac{iot} inference is reflected in multiple aspects, including hardware capabilities, energy resources, and latency tolerance.  To pinpoint some of the core challenges and  motivate the need for different forms of collaboration, as reviewed in the following sections, we next focus on diversity that arises from different levels of mobility of \ac{iot} devices. To highlight this, we consider two extreme settings, as illustrated in Fig.~\ref{fig:DynVsStatic1}:
\begin{itemize}
    \item {\bf Static settings:} here, the \ac{iot} devices are static, and communicate with a cloud server, possibly with the aid of an intermediate access point or edge server, via reliable links of fixed capacity. Such settings include, e.g., the deployment of surveillance cameras in smart cities, wired biomedical monitors in hospitals, or sensor networks in industrial systems. 
    \item {\bf Dynamic settings:} in many applications, \ac{iot} devices are mobile, and are required to operate and infer while traversing through different environments with varying levels of connectivity to the cloud.  For instance, wearable devices, ranging from portable bio-medical sensors to augmented reality systems, should operate also in rural settings without guaranteed access to a centralized server.
\end{itemize}

The above settings represent the extremes of a spectrum of use-cases varying in mobility, with various cases lying in between. The above settings, which represent scenarios where an observation taken at an edge device is to be used for inference, characterize the different requirements of \ac{iot} \ac{dnn}-aided inference.  In the static setting, one can reliably process the data on the cloud server; the main challenges here stem from latency and privacy requirements, as well as the storage and computation capacity of the server, which is much larger compared with edge devices, but also has its limits. These can be tackled via collaborative inference, by properly dividing some of the computation between the edge device and the cloud (and/or intermediate servers in the communication network). In the dynamic setting, connectivity to the cloud is not guaranteed, and inference should be carried out on the edge, where one must cope with the limited ability of edge devices to apply highly parameterized \acp{dnn}. Collaboration in such cases is feasible between different edge devices, communicating in an adaptive device-to-device fashion, with the aim of improving inference accuracy.

The division into static and dynamic settings highlights the need for collaboration, for both remote and on-device inference. These can be related to the main paradigms of cloud computing, fog computing, and edge computing \cite{mahmud2018fog}. The latter infers on the device that acquires the observations, while the former two consider remote inference: in cloud computing, inference is carried out solely on the centralized cloud server, while in fog computing some of the processing is offloaded to intermediate nodes in the communication network, e.g., access points and edge servers. Since in this article we focus on signal processing and algorithmic collaboration approaches, rather than on the design and exploitation of the hierarchical structure of communication networks, we henceforth simplify our categorization into $1)$ {\em cloud-centric inference}, which is carried out remotely regardless of whether it is partitioned between the cloud server and fog nodes; and $2)$ {\em edge inference} that is done on the edge device itself.  

\section{Cloud-Centric Inference}
\label{sec:Cloud}

\begin{figure*}
    \centering
    \includegraphics[width=0.75\linewidth]{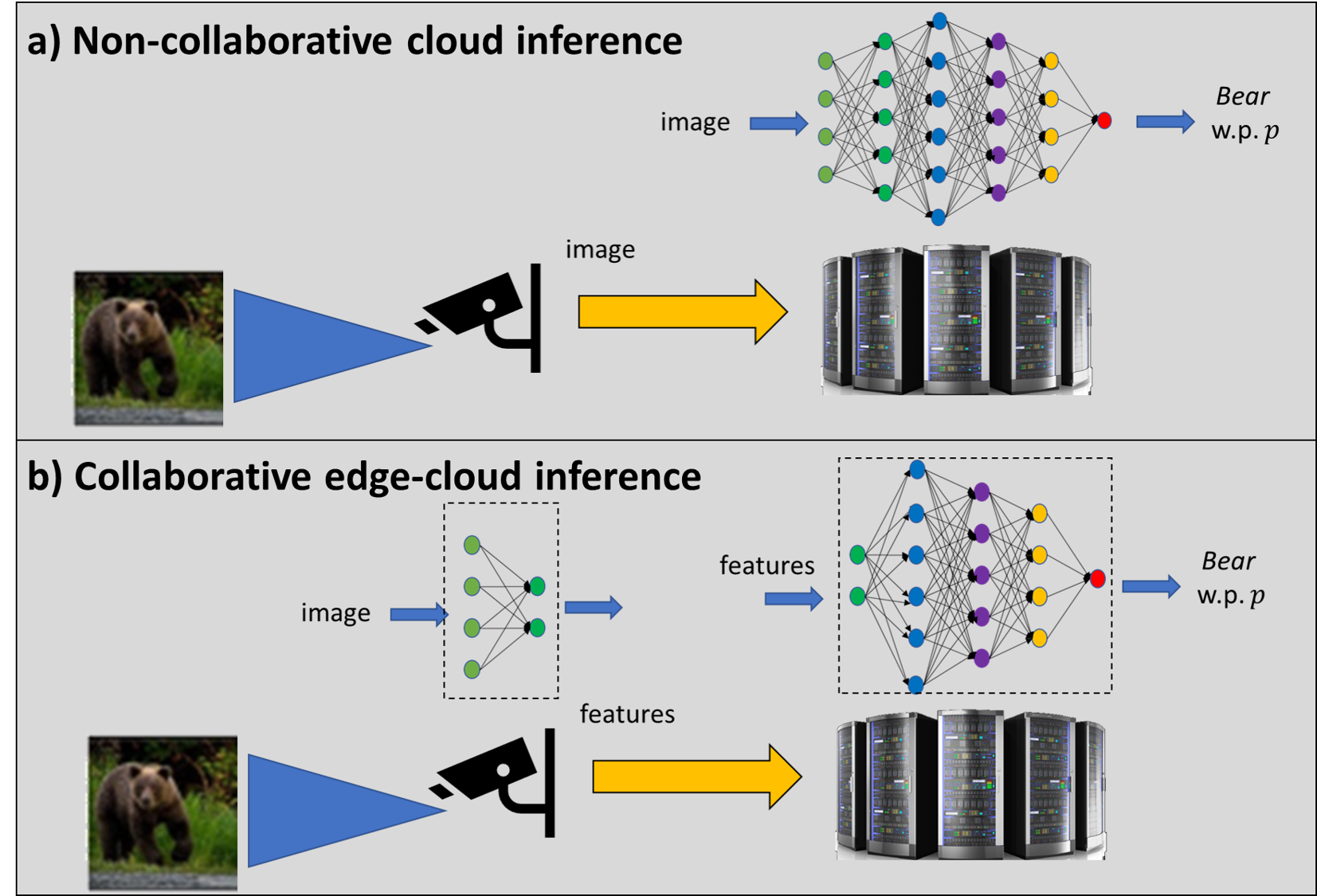}
    \caption{Illustration of existing approaches for \ac{dnn}-aided inference on the cloud of samples acquired by edge devices, for an example scenario of classification of images gathered by a surveillance camera. The depicted schemes include $a)$ non-collaborative cloud inference, where the server processes the raw sample using a highly-parameterized \ac{dnn}; $b)$ collaborative inference by partitioning a \ac{dnn} into a light-weight encoder employed at the edge and a highly-parameterized decoder utilized by the server. }
    \label{fig:InfOnCloud}
\end{figure*}

By \textit{cloud-centric inference} we mean a scenario in which the inference result is produced in the cloud. This result may then be sent back to the edge, if necessary. Two versions of cloud-centric inference are illustrated in Fig.~\ref{fig:InfOnCloud}; in one case (Fig.~\ref{fig:InfOnCloud}(a)), input data, such as an image, is uploaded to the cloud, and in the other case (Fig.~\ref{fig:InfOnCloud}(b)), features derived from the input data are sent to the cloud. Imperfections in the uplink channel will affect the performance in either case. To deal with such channel impairments, for the former strategy, established error resilient coding techniques 
can be used. For the latter one, there has been recent work on decoder-side error mitigation~\cite{bajic_icc2021}, as well as joint source-channel coding~\cite{jankowski2020wireless}.

\subsection{Non-Collaborative Cloud Inference}
\label{subsec:CloudFull}
In non-collaborative cloud inference (Fig.~\ref{fig:InfOnCloud}(a)), the entire DNN is deployed in the cloud. The edge device merely captures the data, for example an image, and sends it to the cloud. An advantage of such a simple scheme is that it is relatively easy to deploy with the current technology -- modern cameras are equipped with advanced hardware-based image/video codecs and communication capabilities, while large DNNs easily run in the cloud. There are, however, a number of downsides to this scheme. First, sending the entire datum (e.g., an image) to the cloud to perform inference uses up more bits than necessary. It is shown in~\cite{cb2022} that, at a given inference accuracy, features from any layer of an arbitrary non-generative DNN are more compressible than its input. This means that uploading intermediate features from a DNN, rather than its input, is more bit-efficient. Reducing the number of uploaded bits could also reduce the overall latency. Finally, uploading data directly to the cloud raises privacy concerns. All these issues may be alleviated via collaborative edge-cloud inference, which we discuss next.

\subsection{Collaborative Edge-Cloud Inference}
\label{subsec:CloudColl}
In this scenario, shown in Fig.~\ref{fig:InfOnCloud}(b), a DNN is partitioned into a front-end (initial few layers) deployed on an edge device, and a back-end (remaining part) residing in the cloud. The edge device computes the features that are then uploaded to the cloud. As discussed above, this strategy is more bit-efficient, which will in turn reduce the communication latency of uploading to the cloud, compared to the non-collaborative cloud inference. The overall inference latency is a combination of this communication latency, the computation time on the edge device, and the computation time in the cloud. 

\begin{figure}
    \centering
    \includegraphics[width=\columnwidth]{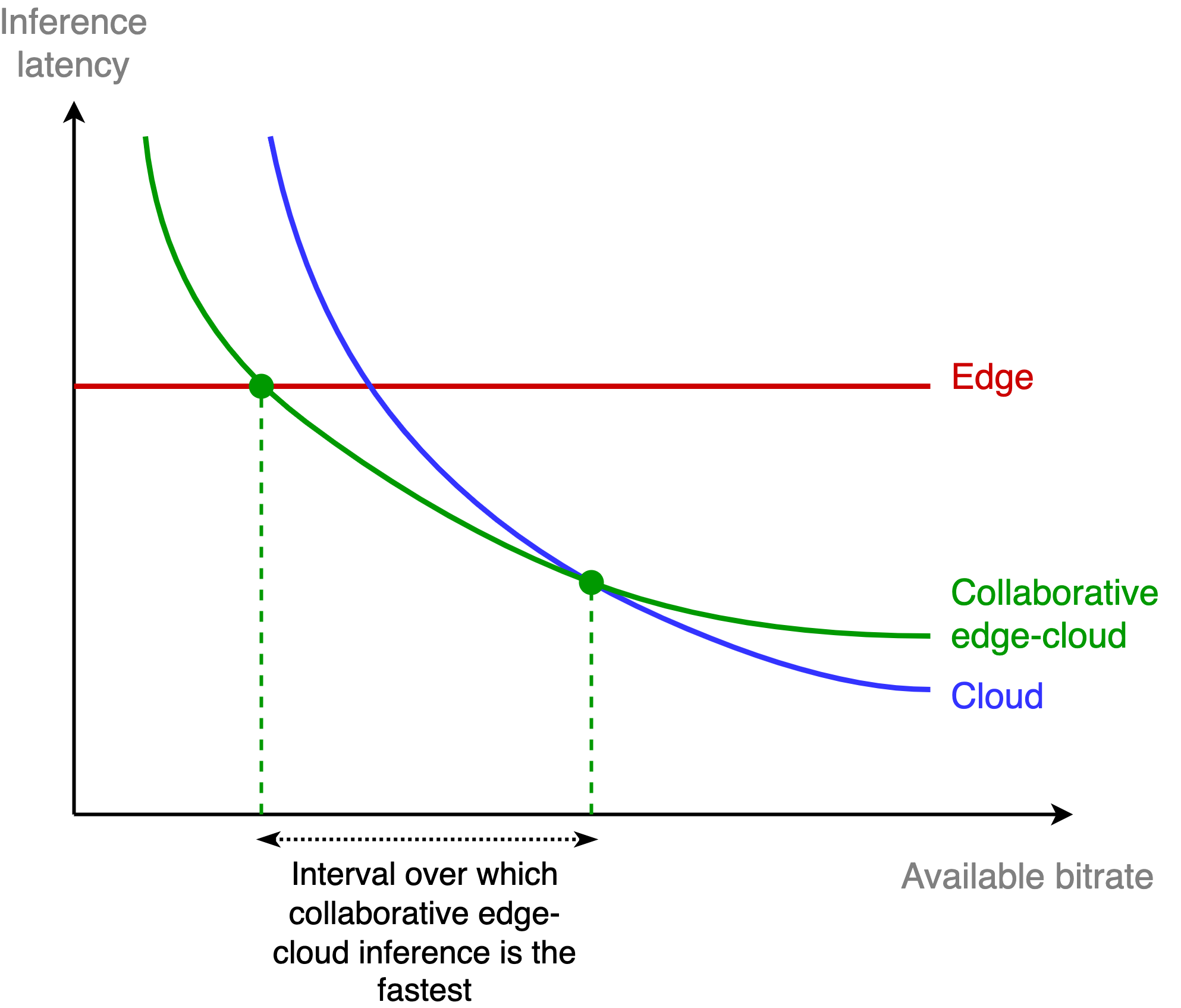}
    \caption{Typical inference latency plots for edge, cloud, and collaborative edge-cloud inference, as a function of available bitrate of the channel connecting the edge device to the cloud.}
    \label{fig:latency}
\end{figure}

Typical plots of the overall inference latency as a function of the available upload bitrate for edge inference (where the entire DNN is on the edge device, see Subsection~\ref{subsec:EdgeFull}), non-collaborative cloud inference, and collaborative edge-cloud inference are shown in Fig.~\ref{fig:latency}, with some actual measurements provided in~\cite{kang2017neurosurgeon}.
In interpreting these plots, it is important to remember that the hardware available in the cloud is faster than that available on the edge device, and that uploading intermediate DNN features is more bit-efficient than uploading its input. When the available bitrate is very small, communication latency is higher than the computation time even on the edge device, and in this regime, edge inference is the fastest. Since edge inference does not require upload to the cloud, its inference latency is shown as a constant. At the other extreme, when the bitrate is large, communication latency becomes negligible compared with the computation time. In this case, non-collaborative cloud inference is the fastest, since the cloud operates faster hardware. In between these extremes, there is an interval of available bitrates over which edge-cloud collaborative inference is the fastest. 

Finally, the fact that collaborative edge-cloud inference avoids uploading original data and sends features instead may alleviate privacy issues. However, features uploaded to the cloud can still carry some private information, which may be revealed through various attacks. 
For example, the \textit{model inversion attack}~\cite{mia_2019} 
tries to recover the original data from the features, which essentially means inverting the front-end of the DNN on the edge device. If successful, private information in the original data will be revealed. There are currently limited defenses against such attacks, one being an information-theoretic approach coined  \textit{privacy fan}~\cite{ab_vcip2021}, where non-private inference-relevant features are lightly compressed while privacy-revealing features are more heavily compressed to remove private information. It should be mentioned that this is a relatively unexplored area, where much future work is required to develop effective solutions, as discussed in Section~\ref{sec:RoadAhead}.



\section{Edge Inference}
\label{sec:Edge}
In this section we consider settings where inference is to be carried out on the edge, e.g., by an \ac{iot} device. Applying \acp{dnn} on \ac{iot} devices as a form of edge computing allows to infer on the same device where the data is collected, rather than having the samples sent to a centralized cloud server. Such \ac{dnn}-aided edge devices can operate at  various connectivity conditions with reduced latency, as well as alleviate privacy issues and facilitate the personalization of  \ac{ai} systems \cite{mao2017survey}.

\begin{figure*}
    \centering
    \includegraphics[width=\linewidth]{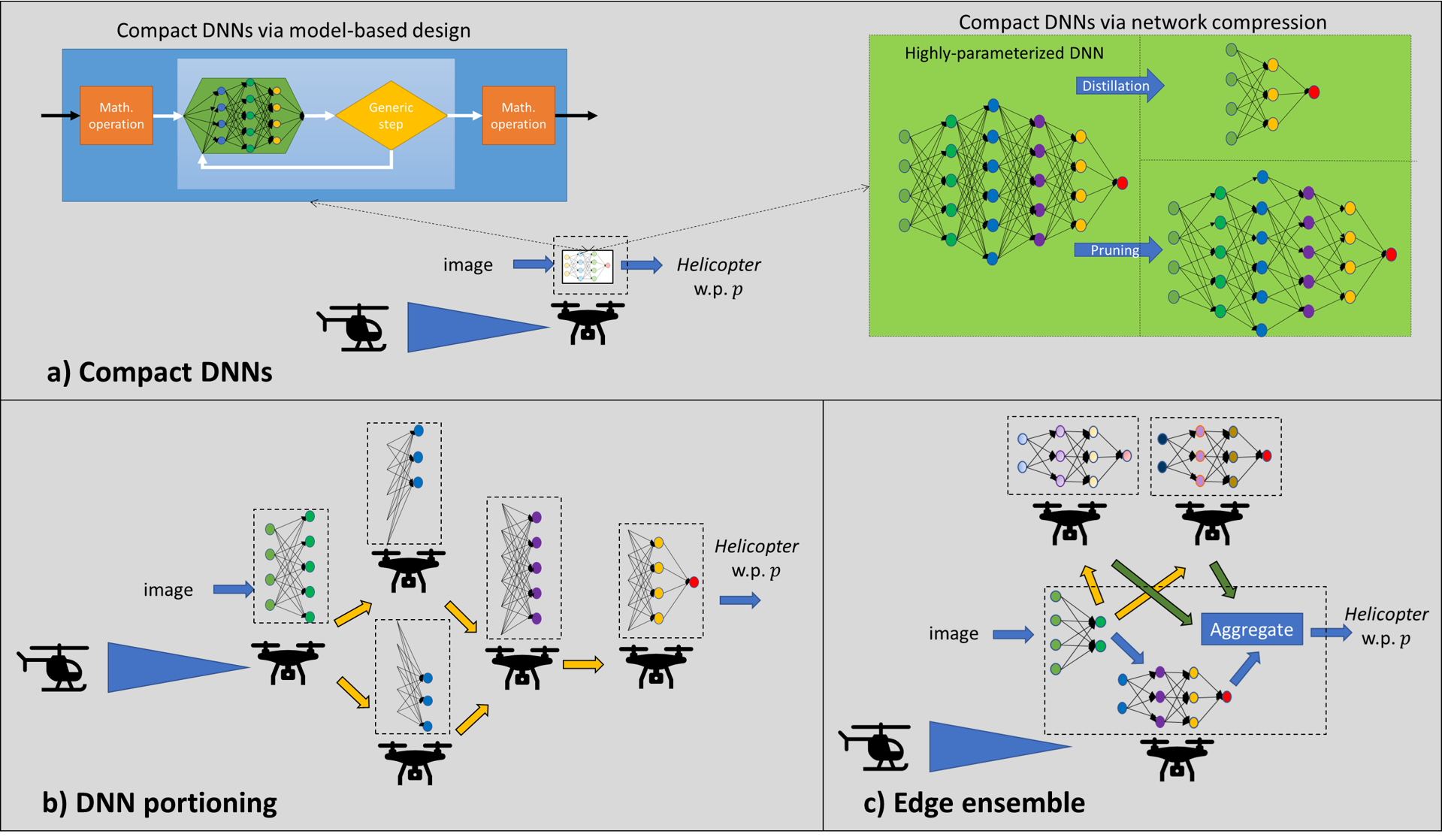}
    \caption{Illustration of existing approaches for \ac{dnn}-aided inference on edge devices, for an example scenario of inference carried out by UAVs. The depicted schemes include $a)$ non-collaborative local inference using compact \acp{dnn}, design via model compression (e.g., distillation or pruning) or via model-aware design; $b)$ collaborative inference by partitioning of a highly-parameterized \ac{dnn} among multiple devices; $c)$ collaborative inference via edge ensembles with diverse \acp{dnn} used by the different collaborating devices. }
    \label{fig:InfOnEdge}
\end{figure*}

The core challenge with applying trained \acp{dnn} on \ac{iot} devices stems from their limited computational resources. 
Existing strategies for on-device \ac{ai}-based inference can be divided into non-collaborative and collaborative ones. 
The former
aims at designing compact \acp{dnn} that are applicable on hardware-limited devices, as illustrated in Fig.~\ref{fig:InfOnEdge}(a) and briefly discussed in Subsection~\ref{subsec:EdgeFull}. Collaborative approaches, which are the focus of this article, leverage the ability of \ac{iot} devices to communicate with neighbouring peers to enable high performance inference. This is achieved via partitioning of complex computations over multiple devices (Fig.~\ref{fig:InfOnEdge}(b)), or by forming an ad hoc deep ensemble (Fig.~\ref{fig:InfOnEdge}(c)). In the following we elaborate on these families of methods. 


\subsection{Non-Collaborative Edge Inference}
\label{subsec:EdgeFull}
To enable the application of trained \acp{dnn} on edge devices without having multiple devices collaborate, one typically has to utilize compact \acp{dnn}. As deep learning usually employs highly-parameterized models, a key challenge here is to design models that are compact without compromising too much on performance. 

Various techniques have been proposed in the literature for compacting \acp{dnn}, see survey \cite{zhang2021compacting}. The conventional framework deals with scenarios where one has access to a highly-parameterized high-performance \ac{dnn}, and aims at making it more compact. Among the leading techniques for compacting \acp{dnn}  are 
{\em knowledge distillation}, where a compact \ac{dnn} is trained to imitate the highly-parameterized pre-trained model; 
{\em network pruning}, which intentionally throws away neurons and/or nullifies weights in the trained model; 
and {\em network quantization}, where the weights of the network are coarsely discretized, possibly to a single bit. 

The above approaches aim at compacting a given \ac{dnn}. Alternatively, one can design an \ac{ai} model to be light-weight in the first place, rather than starting from a pre-trained highly-parameterized \ac{dnn}. For instance,  the fact that a \ac{dnn} is to be pruned or quantized can be accounted for in its training procedure, boosting the trained model to facilitate compaction. 
Furthermore, one can prefer deep architectures, such as convolutional networks with small kernels and shortcut connections \cite{zhang2021compacting}, that are inherently more compact compared with conventional ones. An alternative strategy is to design \ac{dnn}-aided systems that utilize compact networks by incorporating statistical model-based domain knowledge and augmenting a suitable classic inference algorithm with trainable models, see, e.g., survey in \cite{shlezinger2020model}.

Non-collaborative edge-inference strategies focus on a single edge user, e.g., a single \ac{iot} device. As such, they do not exploit the fact that while each device is limited in its hardware, multiple users can confidently collaborate, even in the absence of reliable connectivity to a centralized server. Such collaboration, discussed in the following sections, allows the system  to benefit from the joint computational resources of multiple \ac{iot} devices. 

\subsection{Computation Partitioning}
\label{subsec:EdgeOffload}
The ability of edge devices to communicate and collaborate can be harnessed to enable \ac{ai}-aided inference by partitioning and dividing a highly-parameterized \ac{dnn}  among multiple devices.
Such techniques, coined {\em computation partitioning} or {\em offloading} \cite{mao2017survey}, are schematically illustrated in Fig.~\ref{fig:InfOnEdge}(b). 

There are several schemes to partition a \ac{dnn} among multiple devices. The most straight-forward approach divides the \ac{dnn} by layers (or blocks of layers). In such {\em layer-based} collaborative inference, each participating device only applies a subset of the layers of the \ac{dnn}, and communicates its output features, which are possibly compressed to reduce the overhead, 
to the specific device that applies the subsequent layers. Layer-based partitioning of a \ac{dnn} can also be combined with {\em horizontal} partitioning, where the computations of each layer are divided among multiple users, i.e., different users apply different neurons of the same layer \cite{huang2022toward}. 
The latter is essential when utilizing wide \acp{dnn}, in which some layers may be comprised of too many neurons to be applicable on hardware-limited edge devices.

The partitioning of a multi-layered \ac{dnn} among multiple users  allows to jointly form a large network during inference. As such, in the absence of communication errors, it enables \ac{ai}-aided edge inference without compromising on accuracy compared with cloud-centric inference. Nonetheless,  each user cannot infer on its own, and must rely on the availability of neighbouring nodes, which have access to the required \ac{dnn} partitions. This notably complicates the ability to form an ad hoc collaboration, and typically involves some centralized orchestration and model distribution.  Finally, the repeated communications among the multiple devices, and the potential presence of stragglers due to the heterogeneous nature of \ac{iot} devices, results in possibly increased inference latency, and requires dedicated optimization of the workload and communication among the participating devices~\cite{huang2022toward}. 

\subsection{Edge Ensembles}
\label{subsec:EdgeEnsembles}
The above edge inference approaches either rely solely on local inference (via compact \acp{dnn}), or require reliable communications with a set of edge devices (in computation offloading), which is either pre-defined or determined via a dedicated protocol with additional overhead and orchestration.
Edge ensembles \cite{malka2022decentralized,yilmaz2022over} is a collaborative inference strategy that supports purely local inference while enabling the benefit from collaboration among multiple devices. 

The rationale here is to have the edge devices utilize compact \acp{dnn}, such that each device can infer locally in the absence of reliable connectivity~\cite{malka2022decentralized}. Nonetheless, while the individual models used by each device are designed for the same inference task, they are diverse. For instance, each device may posses a \ac{dnn} with different weights, activations, or architectures. The diversity among the users allows multiple devices to form a deep ensemble during inference, by aggregating predictions made by multiple devices. Doing so leverages the known performance gains of ensemble models, allowing devices to improve inference accuracy via collaboration. 

The direct scheme for edge ensembles requires the inferring user to share the observations with its available neighbouring devices. Each  participating user applies its local \ac{dnn} to the data and transmits the prediction to the inferring user, which in turn aggregates them into a decision via, for instance, ensemble averaging, majority vote, or even as a form of over-the-air computation \cite{yilmaz2022over}. As sharing the observations, e.g., a set of images, may give rise to communication latency and privacy concerns, one can also first apply a shared encoder to map the data into low-dimensional features that are shared with the neighbours, which use diverse decoders for local inference. This procedure is illustrated in Fig.~\ref{fig:InfOnEdge}(c).   

\begin{figure}
    \centering
    \includegraphics[width=\columnwidth]{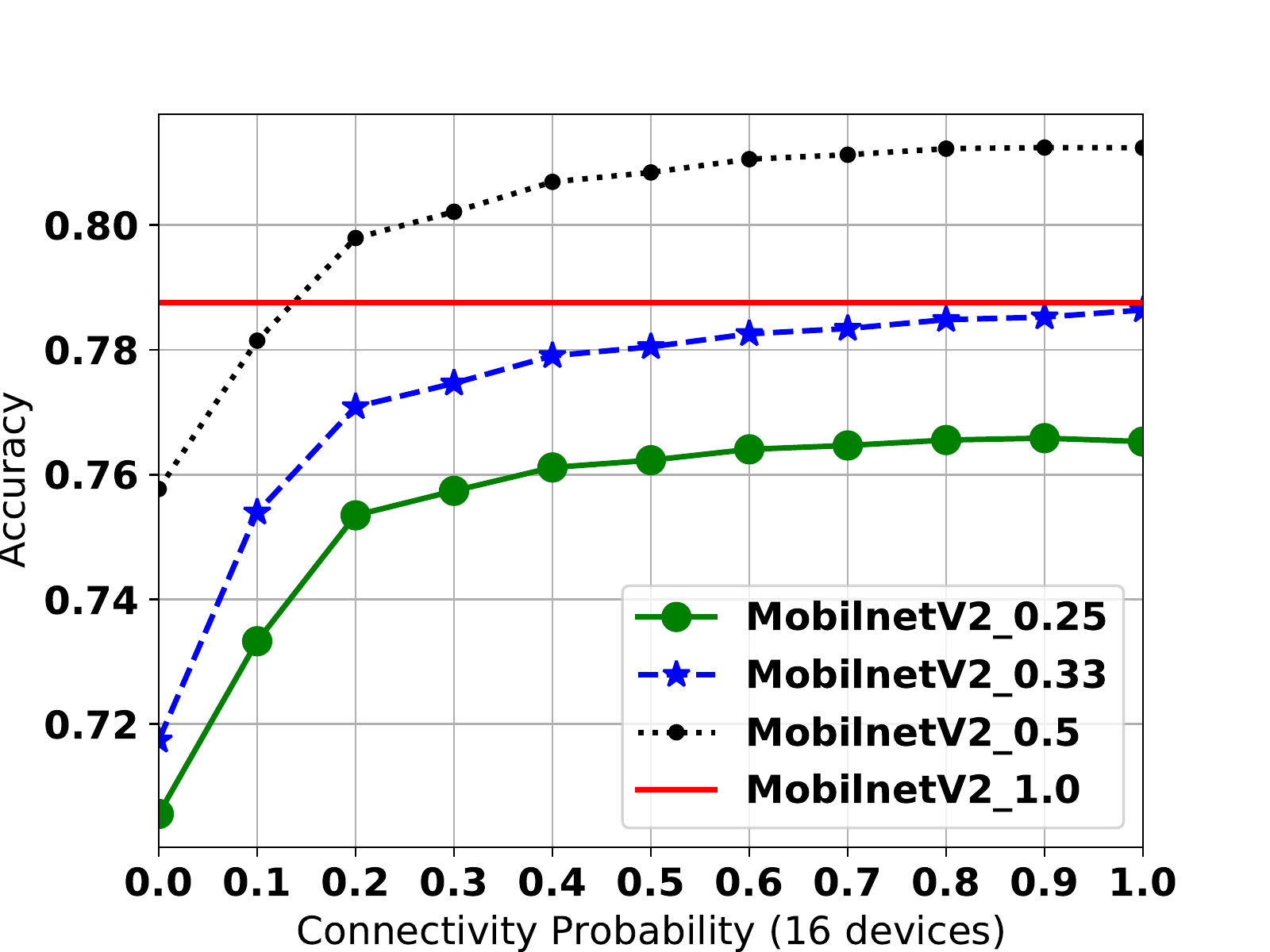}
    \caption{Accuracy comparison of edge ensembles with up to $16$ users when inferring with MobilenetV2 with different width factors compared with centralized inference with a large network of width factor $1$.}
    \label{fig:p_models_16}
\end{figure}

Edge ensembles are inherently adaptive. 
When the only participating user is the one inferring, e.g., in the absence of available neighbouring devices, edge ensemble boils down to local inference discussed in Subsection~\ref{subsec:EdgeFull}. Collaboration can be formed ad hoc in a decentralized manner, making it suitable for mobile settings with varying connectivity levels. While each user infers with a compact \ac{dnn} whose performance may be limited, collaboration mitigates the performance loss compared to using a single large \ac{dnn}, and in fact allows to achieve performance improvements in some settings. This is exemplified in Fig.~\ref{fig:p_models_16}, which depicts the accuracy achieved by an edge ensemble for image classification with the CIFAR10 data set versus the connectivity probability for 16 devices. 
The edge devices infer with diverse MobilenetV2 \acp{dnn} with width factors of $\frac{1}{4}, \frac{1}{3}$, and $\frac{1}{2}$, for which the number of parameters is  $\{2.51, 3.96,7\}\cdot10^5$, respectively, and the accuracy is compared with a baseline centralized full MobilnetV2 (abbreviated {\em MobilenetV2\_1.0}) comprised of over $2.3\cdot 10^6$ parameters. The complete details of the numerical study can be found in \cite{malka2022decentralized}. It is observed in Fig.~\ref{fig:p_models_16} that while the full MobilenetV2 is the most accurate among the individual models, collaboration among multiple users allows devices utilizing a \ac{dnn} with width factors of $\frac{1}{4}$ and $\frac{1}{2}$, which have roughly $6\%$ and $30\%$ the number of weights as MobilenetV2\_1.0, respectively, to approach and even outperform the large centralized model.

\section{Summary and the Road Ahead}
\label{sec:RoadAhead}
Here, we provide a qualitative comparison between the aforementioned strategies for \ac{dnn}-aided edge inference. Then,  we detail a few important research directions that should be explored to fully unveil the potential of the considered strategies, and provide concluding remarks.

\begin{table*}
\begin{tabular}{|p{1.2cm}|p{1.7cm}|p{1.8cm}|p{2.9cm}|p{2.5cm}|p{2.5cm}|p{2.5cm}|}
 \hline
                                & {\bf Method}                           & {\bf Collaborate}        & {\bf Accuracy}            & {\bf Communications}                                                & {\bf Privacy}                                                             & {\bf Connectivity}                                        \\
                                \hline \hline
\multirow{2}{1.2cm}{Cloud-centric inference}  & Cloud inference                  & None                 & {\em Highest} -- usage of large DNNs                                              & High -- due to communications of observations                    & None -- data shared over communications network              & Requires reliable high-throughput link to server    \\   \cline{2-7}
                                & Collaborate edge-cloud inference & Edge-cloud           & High -- usage of large DNNs, possible distortion due to feature compression & Medium - sharing of compressed features                              & Partial - shared features that can be crafted to enhance privacy & Requires reliable link server                       \\  \hline
\multirow{3}{1.2cm}{Edge inference} & Compact networks                 & None                 & Typically degraded due to network compacting                                                 & {\em Minimal} -- no communications                                          & {\em Fully private} - no data sharing                                     & {\em Invariant}                                           \\ \cline{2-7}
                                & Computation partitioning         & Multiple edge devices & High -- usage of large DNNs, possible distortion due to feature compression                                              & High -- due to repeated device-to-device communications               & Partial -- shared features that can be crafted to enhance privacy & Requires reliable links with specific edge devices \\   \cline{2-7}
                                & Edge ensembles                   & Multiple edge devices & Adaptive -- degraded in low connectivity, increases when collaboration is feasible        & Low -- multi-casting of compressed features via device-to-device links & Partial -- shared features that can be crafted to enhance privacy & {\em Fully adaptive} -- operable in different connectivity levels \\  \hline
\end{tabular}
    \caption{Qualitative comparison between the considered approaches for \ac{iot} \ac{ai}-empowered inference. }
    \label{Tbl:Comparison}
\end{table*}

\subsection{Comparison}
\label{subsec:summary}
The approaches detailed in Sections~\ref{sec:Cloud}-\ref{sec:Edge} for facilitating \ac{ai}-empowered \ac{iot} inference differ in their properties, and are each suitable for different types of scenarios. Broadly speaking, cloud-centric inference is geared towards settings with guaranteed reliable connectivity, e.g., static \ac{iot} systems, while edge inference is most suitable for applications with strict latency constraints or a need to operate in mobile settings. To provide a meaningful  comparison, we focus on four key figures-of-merit -- inference accuracy, communication latency, privacy, and connectivity requirements. 

\subsubsection{Accuracy} The starting point for \ac{ai}-empowered inference is typically some large high-performance \ac{dnn}, whose accuracy is degraded by modifying and compressing it. As such, inferring solely on the cloud, which can host the large \ac{dnn}, is expected to be most accurate.  Techniques that partition the \ac{dnn}, either via collaborative edge-cloud inference or by offloading over multiple edge devices, may induce some degradation as features being shared between the entities often undergo lossy compression. Compacting the \ac{dnn} is likely to yield the most notable degradation, though its effect can be mitigated by collaboration via edge ensembles. 

\subsubsection{Communications} Non-collaborative inference on the edge does not entail any communication overhead, as processing is carried out solely on-device, and thus offers the least communication latency of all considered methods. Cloud-centric inference may involve notable communication latency due to the need to convey the observed data from the edge to the cloud over the network, though this overhead can be reduced by sharing compressed features via collaborative edge-cloud inference. Among collaborative edge inference schemes, computation partitioning may induce substantial communication latency due to the repeated exchange of features among the participating devices and the need to coordinate the procedure, while edge ensembles entails minimal excessive overhead, as it involves a single round of multi-casting compressed features between the inferring user and its neighbouring devices. 

\subsubsection{Privacy} The data used for inference may contain private information. Thus, sharing the data over the communication network, as done in cloud inference, does not preserve privacy. In all collaborative schemes, one can enhance privacy by sharing extracted features rather than the data itself, though this requires dedicated crafting of the features. Clearly, inferring locally with a compact \ac{dnn} is most privacy preserving. 

\subsubsection{Connectivity} Cloud-centric inference requires reliable connectivity between the \ac{iot} device and the cloud server, where non-collaborative cloud inference needs high-throughput links for sharing the full observations. Edge inference can typically be robust to limited connectivity and applicable in mobile settings, though computation partitioning still requires reliable communications between a (possibly pre-defined) set of users that jointly possess all the partitions of a complete large \ac{dnn}.

The comparison detailed above is summarized in Table~\ref{Tbl:Comparison}.

\subsection{Future Research Directions}
\label{subsec:directions}
Collaborative inference bears the potential of paving the way to a smooth transition of \ac{ai} from the domain of powerful centralized servers, into a multitude of easily accessible and portable devices. However, there are several research directions that should be further explored in order to realize the potential of these methods. In the following we discuss a few representative topics, considering both theoretical studies as well as algorithmic aspects and system design. 

\subsubsection{Privacy Guarantees} 
Collaboration during inference naturally gives rise to privacy considerations, as it involves sharing of data samples that may contain private information. However, measuring privacy, which is essential to characterizing privacy guarantees, is not trivial in the context of inference tasks. A widely-accepted concept in the machine learning literature is differential privacy, which deals with guaranteed obscuring of individual samples in large data sets, and may thus be less suitable for inference based on a single data sample. An alternative privacy measure is information theoretic privacy considered in \cite{ab_vcip2021}, which represents the statistical dependence between the shared and private features. The study and characterization of meaningful privacy measures for inference tasks is thus critical for allowing collaborative inference that is free of privacy concerns.  

\subsubsection{Hybrid Collaboration Design} 
The division into cloud-centric and edge inference is motivated by the categorization of static and mobile \ac{iot} settings discussed in Section~\ref{sec:Inference}. However, one can also expect situations involving both static and mobile users. For instance, inference carried out by a mobile autonomous vehicle that can ad hoc communicate with both neighbouring vehicles and road-side units that are wired to the network infrastructure. Such scenarios motivate the study of hybrid collaboration strategies, which can adaptively benefit from collaboration among multiple edge devices, as in edge ensembles, while leveraging possible connectivity with a centralized server, as in collaborative edge-cloud inference. 

\subsubsection{Over-the-Air Computations} 
Recent years have witnessed a growing interest in  over-the-air computation techniques for reducing the communication latency when learning on the edge, particularly in a federated manner \cite{gafni2021federated}. These techniques are most relevant when the edge users communicate over a shared wireless channel, such that a desired joint computation can be achieved by non-orthogonal synchronized communications with suitable precoding. This motivates studying over-the-air schemes for edge inference, initially explored in \cite{yilmaz2022over}, being naturally suitable for edge ensembles, with the promise of reducing latency and possibly enhancing privacy (as the channel noise is now added to the shared features and decisions).

\subsubsection{Joint Hardware-Algorithmic Design}
While we focus on algorithmic aspects to facilitate \ac{ai}-empowered edge inference, using \acp{dnn} on \ac{iot} devices will also involve hardware developments. The fact that \ac{dnn}-aided edge inference will require novelty in both system hardware and collaboration algorithms indicates the potential of joint hardware-algorithmic designs. For instance, algorithmic approaches can possibly guide hardware design, or alternatively, the characteristics of hardware accelerators can reveal unique requirements for efficient collaborative inference. 

\subsubsection{Diverse Edge Models} 
Collaboration via edge ensembles requires having diverse models among the users. Achieving such diversity is thus key to this form of collaboration. When training is done on the edge, diverse models typically emerge as the training data differs between devices. Diversity can also be obtained when compacting a trained large \ac{dnn} by, e.g., having each small model trained with a different initialization via knowledge distillation, or by compressing a large network with stochastic quantization, resulting in different compressed realizations of the same network. These initial ideas, though, all require further exploration to understand what technique is most suitable for which scenario.

\subsection{Conclusions}
\label{subsec:conclusions}
In this article, we reviewed approaches for facilitating \ac{ai}-aided \ac{iot} inference via collaboration. We categorized existing methods into two main strategies -- cloud-centric inference and edge inference -- and highlighted the main characteristics and challenges of each approach. 
By harnessing collaboration, either with a cloud server or among multiple devices, it is shown that one can achieve improvements compared with inferring solely on the cloud and/or an edge device in terms of accuracy, communication latency, privacy, and adaptivity. We discussed several research directions that arise from the concept of collaborative inference, which are expected to pave the way in unveiling its potential for bringing \ac{ai} to the edge.

\bibliographystyle{IEEEtran}
\bibliography{IEEEabrv,refs}

\end{document}